\documentclass{PoS}
\usepackage{epsfig,amsmath,amsfonts,amssymb,amstext,afterpage,psfrag,slashed}
\usepackage{longtable}

\newcommand{\ewXL}{EWCh$\mathcal{L}$}

\DeclareMathOperator{\tr}{tr}
\DeclareMathOperator{\str}{str}
\DeclareMathOperator{\Str}{Str}

\title{Complete One-Loop Renormalization of the Higgs-Electroweak Chiral Lagrangian}

\ShortTitle{1-loop Renormalization of the \ewXL}

\author{\speaker{Claudius Krause}\thanks{CK acknowledges the support of the Alexander von Humboldt Foundation. }\\
        Theoretical Physics Department, Fermi National Accelerator Laboratory, Batavia, IL,60510, USA\\
        E-mail: \email{ckrause@fnal.gov}}

\author{Gerhard Buchalla\\
        Ludwig-Maximilians-Universit\"at M\"unchen, Fakult\"at f\"ur Physik, Arnold Sommerfeld Center for Theoretical Physics, D-80333 M\"unchen, Germany\\
        E-mail: \email{Gerhard.Buchalla@physik.uni-muenchen.de}}

\author{Oscar Cat\`a\\
Theoretische Physik 1, Universit\"at Siegen, Walter-Flex-Stra\ss e 3, D-57068 Siegen, Germany\\
        E-mail: \email{Oscar.Cata@uni-siegen.de}}

\author{Alejandro Celis\\
        Ludwig-Maximilians-Universit\"at M\"unchen, Fakult\"at f\"ur Physik, Arnold Sommerfeld Center for Theoretical Physics, D-80333 M\"unchen, Germany\\ 
        E-mail: \email{alejandrocelis5555@gmail.com}}

\author{Marc Knecht\\
        Centre de Physique Th\'eorique (CPT),UMR 7332 CNRS/Aix-Marseille Univ./Univ. du Sud Toulon-Var, Marseille, France\\
        E-mail: \email{marc.knecht@cpt.univ-mrs.fr}}

\abstract{The electroweak sector of the Standard Model can be formulated in a way similar to Chiral Perturbation Theory (ChPT), but extended by a singlet scalar. The resulting effective field theory (EFT) is called Higgs-Electroweak Chiral Lagrangian (\ewXL) and is the most general approach to new physics in the Higgs sector. It solely assumes the pattern of symmetry breaking leading to the three electroweak Goldstone bosons ({\it i.e.} massive $W$ and $Z$) and the existence of a Higgs-like scalar particle. The power counting of the \ewXL\ is given by a generalization of the momentum expansion of ChPT. It is connected to a loop expansion, making the theory renormalizable order by order in the EFT.\\
  I will briefly review the construction of the \ewXL\ and its power counting. Then, I will discuss the complete one-loop renormalization of the \ewXL\ employing the background-field method and the super-heat-kernel expansion. This computation confirms the power counting assumptions, is consistent with the completeness of the operator basis, and reproduces known results of subsectors in the appropriate limits.\\
Preprint numbers: FERMILAB-CONF-19-066-T}

\FullConference{The 9th International workshop on Chiral Dynamics\\
		17-21 September 2018\\
		Durham, NC, USA}

\begin{document}

\section{Introduction and Motivation}
After the discovery of a Higgs-like scalar particle at the LHC, the particle physics community was interested in answering the question\\
{\it ``Is that particle the Higgs-boson of the Standard Model, or is it something else?''}  \\
Many models of new physics beyond the Standard Model (SM) exist, and many of them predict new particles at the scales accessible to the LHC. So far, no such direct sign of new physics has been discovered and the focus of the analyses shifted towards searches for indirect effects. In those searches, the couplings of the SM particles are measured with high precision and deviations from the SM expectations are looked for. \\
Bottom-up effective field theories (EFTs) offer the perfect tool for these analyses, as they are consistent quantum field theories while at the same time they only assume very little on the new physics. In the bottom-up approach, the couplings of the effective operators (called Wilson coefficients) are free parameters to be determined by experiment. Once a UV-model is specified, the Wilson coefficients can be computed and the EFT can be specified to the given model. With all Wilson coefficients kept as free parameters, the bottom-up EFT provides a model-independent approach to look for the indirect signs of new physics.\\
Since the couplings of the Higgs-like scalar are so far only known with $\mathcal{O}(10\%)$~\cite{deBlas:2018tjm} uncertainty or more, we cannot assume that it is indeed the Higgs of the SM and new physics decouples from the SM. Dynamical symmetry breaking, like in composite Higgs models~\cite{Kaplan:1983fs,Dugan:1984hq} where the Higgs-like scalar is a pseudo-Nambu-Goldstone boson, is still an appealing alternative to solve the hierarchy problem.\\
The Electroweak Chiral Lagrangian (\ewXL) is the most general EFT for the Higgs sector and therefore the most suitable bottom-up EFT for current LHC analyses. Its construction can be understood intuitively as follows. First, we start from the Lagrangian of the SM and omit the physical Higgs scalar. The resulting Higgs-less chiral Lagrangian is then supplemented with a generic scalar singlet with the most general couplings to fermions and gauge fields. The Goldstone dynamics in the chiral Lagrangian is given by the pattern of symmetry breaking $SU(2)_{L}\times SU(2)_{R}\to SU(2)_{L+R}$. This is analogous to QCD and Chiral Perturbation Theory (ChPT). The underlying strongly-coupled UV theory (QCD) has a global chiral symmetry that is dynamically broken at low energies. The dynamics of the resulting low-energy EFT (ChPT) is solely governed by the pattern of this global symmetry breaking. Moreover, the chiral Lagrangian is agnostic about how the electroweak symmetry is broken. It can describe the SM in the limit described below in eq.~\eqref{smlimit}, as well as composite Higgs models, as discussed in~\cite{Buchalla:2014eca,Krause:2018cwe}.\\
Parallel to the \ewXL, there is an alternative EFT used to study new physics beyond the SM, the Standard Model EFT (SMEFT)~\cite{Grzadkowski:2010es}. In the SMEFT, the Higgs belongs to the $SU(2)_{L}$ Higgs doublet of the SM and the effects of decoupling new physics is described by operators of canonical dimension larger than four. \\
With the increased interest in EFTs for data analysis at the LHC, there was also an increased interest in formal aspects of both EFTs, like (among others) the counting of independent operators~\cite{Henning:2015alf} in SMEFT, the matching to models~\cite{Henning:2016lyp,Fuentes-Martin:2016uol,Zhang:2016pja}, a geometric interpretation of the scalar field space~\cite{Alonso:2016oah}, and the one-loop renormalization~\cite{Jenkins:2013zja,Jenkins:2013wua,Alonso:2013hga,Guo:2015isa,Alonso:2017tdy,Buchalla:2017jlu}. \\
In these proceedings, I will report on the one-loop renormalization of the \ewXL~\cite{Buchalla:2017jlu}, which is especially interesting for the following reasons. First, it confirms the power counting that is based on the loop expansion and the superficial degree of divergence~\cite{Buchalla:2013eza}. Second, it provides independent support to the completeness of the basis we previously derived in~\cite{Buchalla:2013rka}. Third, it gives us the running of the couplings. And fourth, the functional approach we employ reduces the problem to an algebraic problem of matrix multiplication and tracing. The resulting master formula can also be applied to other Lagrangians beyond our EFT. 
\section{The Electroweak Chiral Lagrangian}
The \ewXL\ applies the techniques of ChPT to the Higgs sector. As any bottom-up EFT, it is defined by the low-energy particle content, its symmetry content and the power counting of the expansion. 
\begin{itemize}
\item {\bf Particles:} The \ewXL\ assumes the particle content of the SM, including the Higgs-like scalar $h$ and the three EW Goldstone bosons (GBs) $\varphi_{a}$, but no relation between $h$ and the GBs.  
\item {\bf Symmetries:} The \ewXL\ assumes the $SU(3)_{c}\times SU(2)_{L}\times U(1)_{Y}\to SU(3)_{C}\times U(1)_{em}$ local symmetry of the SM and the conservation of baryon number B and lepton number L. It further assumes custodial symmetry at leading order. 
\item {\bf Power counting:} The \ewXL\ is expanded in chiral dimensions~\cite{Buchalla:2013eza}, a generalization of the momentum expansion of ChPT. The total chiral dimension of a term in the Lagrangian sums to $2L+2$, where $L$ is the order of the EFT expansion. Derivatives, fermion bilinears and weak couplings (gauge and Yukawa) have chiral dimension 1, bosons have chiral dimension 0. Leading order is given by $L=0$, {\it i.e.} $\mathcal{O}(p^{2})$ and next-to-leading order by $L=1$, {\it i.e.} $\mathcal{O}(p^{4})$.
\end{itemize}
Once we relax the relation between the $h$ and the GBs, they in general cannot be written as elements of the same multiplet. To have gauge invariance manifest, we introduce polar coordinates and collect the GBs in $U=\exp{(2i\varphi_{a}T_{a}/v)}$, with $v=246~$GeV the electroweak vacuum expectation value and $T_{a}$ the generators of $SU(2)$. A gauge transformation $g_{L}\in SU(2)_{L}$ and $g_{Y}\in U(1)_{Y}$ acts linearly on $U\to g_{L} U g_{Y}^{\dagger}$, but non-linearly on $\varphi_{a}$. Hence the name nonlinear EFT for the \ewXL. The radial mode, $h$, remains a singlet under the symmetry.\\
For generic Higgs couplings, the resulting Lagrangian is non-renormalizable in the traditional sense, having interactions beyond canonical dimension four in the expansion of the exponential in $U$. It is, however, renormalizable in the modern sense --- order by order in the EFT expansion. Counterterms needed for the renormalizatin of loops of leading-order vertices are included as next-to-leading order operators~\cite{Buchalla:2013rka,Buchalla:2013eza}. This ultimately connects the expansion of the EFT to an expansion in loops, which is reflected in the definition of the chiral dimensions above. \\
The leading-order Lagrangian (at chiral dimension 2) is 
\begin{align}
  \begin{aligned}
    \label{eq:l2}
    {\mathcal L} &= -\frac{1}{2} \langle G_{\mu\nu}G^{\mu\nu}\rangle
    -\frac{1}{2}\langle W_{\mu\nu}W^{\mu\nu}\rangle 
    -\frac{1}{4} B_{\mu\nu}B^{\mu\nu}
    +\frac{v^2}{4}\ \langle D_\mu U^{\dagger} D^\mu U\rangle\, F(h) +
    \frac{1}{2} \partial_\mu h \partial^\mu h - V(h)
    \\
    & +\bar\psi i\slashed{D}\psi - \bar\psi \left[U M(h) P_R +M^\dagger(h) U^\dagger P_L\right]\psi.
  \end{aligned}
\end{align}
Here, $G,W,$ and $B$ are the gauge fields of $SU(3)_{C}, SU(2)_{L},$ and $U(1)_{Y}$, respectively; $\langle x \rangle$ denotes the trace of the operator $x$; the covariant derivative of $U$ is defined as
\begin{equation}
  \label{eq:2}
  D_\mu U=\partial_\mu U + i g W_\mu U -i g' B_\mu U T_3\, ,
\end{equation}
with $T_{3}$ the third generator of $SU(2)$; $\psi$ collects the SM fermions, $\psi=(u,d,\nu,e)^T$; and $F(h), V(h),$ and $M(h)$ are Higgs-dependent polynomials containing all powers of $h$. $M(h)$ carries not only fermion-, but also flavor indices.\\
These polynomials generalize the Higgs couplings of the SM. The couplings linear in the Higgs field are related~\cite{Buchalla:2015wfa,deBlas:2018tjm} to the experimental $\kappa$-formalism~\cite{LHCHiggsCrossSectionWorkingGroup:2012nn}. Current LHC data constrains them to be SM-like with roughly $10\%$ uncertainty~\cite{deBlas:2018tjm}. The \ewXL\ is especially suited for the analysis of current Higgs data, because its power counting introduces a hierarchy between the experimentally well-constrained electroweak precision data (at NLO in the \ewXL) and the not-so-well known Higgs couplings (at LO).  

\section{Background-Field Method and Super-Heat-Kernel Expansion}
We employ the background-field method~\cite{Abbott:1981ke} in our computation, splitting each field into a (classical) background component and a (quantized) fluctuating component. We call the latter $\xi, \omega_{\mu},$ and $\chi$ for scalars, vectors, and Dirac fermions, respectively. Further, we define the bosonic objects $\phi_{i}\equiv(\xi,\omega_{\mu})$ and $\phi^{i}\equiv (\xi,-\omega^{\mu})$. Starting from a generic Lagrangian that is at most bilinear in fermion fields and expanding it to second order in fluctuating fields, we have
\begin{equation}
  \label{eq:lfluct}
{\mathcal L}_2= -\frac{1}{2}\phi^i A_i^{\,j} \phi_j + \bar\chi\left( i \slashed{\partial} - G\right) \chi +\bar\chi\Gamma^i\phi_i +\phi^i\bar\Gamma_i\chi,
\end{equation}
with
\begin{equation}
  \label{eq:StFbos}
A = (\partial^\mu + N^\mu) (\partial_\mu + N_\mu) + Y
\end{equation}
and
\begin{equation}
  \label{eq:gdef}
G\equiv (r+ \rho_\mu \gamma^\mu)P_R + (l+ \lambda_\mu \gamma^\mu)P_L.
\end{equation}
The background-field-dependent entities $N_{\mu}, Y, r, \rho_{\mu}, l, \lambda_{\mu}$ are bosonic; $\Gamma^i, \bar\Gamma_i$ are Dirac spinors. The gauge-fixing Lagrangian for $B_{\mu}, W_{\mu},$ and $G_{\mu}$ with fluctuating fields $b_{\mu}, w_{\mu},$ and $\epsilon_{\mu}$ is 
\begin{equation}
  \label{eq:gfix2}
{\mathcal L}_{\rm g.f.} = 
-\frac{1}{2} \left(\partial^\mu b_\mu +\frac{g' v}{2} F\varphi_3\right)^2 
-\frac{1}{2} \left(D^\mu w^a_\mu-\frac{gv}{2} F \varphi^a\right)^2
-\frac{1}{2} \left(D^\mu \epsilon^a_\mu\right)^{2}, 
\end{equation}
where the terms with the fluctuating GBs, $\varphi_{i}$, are introduced~\cite{Dittmaier:1995ee} to cancel a mixing term between the GBs and the gauge fields. This makes the intermediate steps of the computation easier\footnote{We verified explicitly that using a gauge-fixing Lagrangian without these terms leads to the same final result.}. We use the Feynman gauge for the fluctuating gauge fields to obtain the form of eq.~\eqref{eq:StFbos}.\\
The one-loop effective action is given by the Gaussian integral over the bosonic and fermionic fields in
\begin{equation}\label{eq:seffdef}
e^{iS_{\rm eff}} \sim \int [d\phi_i\, d\chi\, d\bar\chi]
e^{i\int d^Dx\, {\cal L}_2(\phi_i,\chi,\bar\chi)}.
\end{equation}
Since we are only interested in the divergent part of $S_{\rm eff}$, we can write~\cite{Neufeld:1998js}
\begin{equation}
  \label{eq:seffstr}
  S_{\rm eff}=\frac{i}{2} \Str\, \ln\Delta,
\end{equation}
where
\begin{equation}
  \label{eq:deltadef}
\Delta\equiv\begin{pmatrix}
A & \sqrt{2}\bar\Gamma\gamma_5 B\gamma_5 \\
-\sqrt{2}\Gamma & B\gamma_5 B\gamma_5
\end{pmatrix}\equiv (\partial^\mu +\Lambda^\mu)(\partial_\mu +\Lambda_\mu)
+\Sigma,
\end{equation}
and $B\equiv i \slashed{\partial} - G $. In this notation, $\Delta$ is a supermatrix treating bosonic and fermionic fields on the same footing. The supertrace of a supermatrix $M=\begin{pmatrix}a & \alpha \\ \beta & b\end{pmatrix}$ is defined\footnote{Capital (super) traces include an integration over space-time, lower-case (super) traces do not include such an integration.} as $\str\, M = \tr\, a - \tr\, b$. On the right hand-side of eq.~\eqref{eq:deltadef}, we brought $\Delta$ already to standard form in super space, defining the super matrices $\Lambda_{\mu}$ and $\Sigma$. The divergent part of the effective action is given by the second Seeley-DeWitt coefficient of the corresponding heat-kernel expansion~\cite{Neufeld:1998js,Donoghue:1992dd}. Using dimensional regularization, the divergent terms of the one-loop effective Lagrangian is given by
\begin{equation}
  \label{eq:ldiv}
{\mathcal L}_{\rm div} =\frac{1}{32\pi^2\varepsilon}\,
\str\, \left[\frac{1}{12}\Lambda^{\mu\nu}\Lambda_{\mu\nu}
+\frac{1}{2}\Sigma^2\right]
\end{equation}
where $\varepsilon = 2 - d/2$ and $\Lambda_{\mu\nu}\equiv \partial_\mu \Lambda_\nu -\partial_\nu \Lambda_\mu+[\Lambda_\mu,\Lambda_\nu]$.\\
With the definitions in eqs.~\eqref{eq:lfluct}--\eqref{eq:gdef}, we can evaluate the supertraces and express the result in terms of the entities $N_{\mu}, Y, r, \rho_{\mu}, l, \lambda_{\mu}, \Gamma^i,$ and $ \bar\Gamma_i$. We can also perform the trace in Dirac space explicitly. We find the master formula~\cite{tHooft:1973bhk,Buchalla:2017jlu}
\begin{align}
  \begin{aligned}
    \label{eq:masterformula}
    {\mathcal L}_{\rm div} = \frac{1}{32\pi^2\varepsilon}\Bigg( &\tr  \Bigg[
    \frac{1}{12} N^{\mu\nu}N_{\mu\nu}+\frac{1}{2} Y^2-\frac{1}{3}\left(\lambda^{\mu\nu}\lambda_{\mu\nu}+\rho^{\mu\nu}\rho_{\mu\nu}\right)    + 2 D^\mu l D_\mu r - 2 lrlr \Bigg] \\ 
    & + \bar\Gamma\left( i\slashed{\partial} +i \slashed{N} +
      \frac{1}{2}\gamma^\mu G\gamma_\mu\right)\Gamma\Bigg)
  \end{aligned}
\end{align}
with
\begin{align}
  \begin{aligned}
    \label{eq:mfdef}
    N_{\mu\nu} \equiv  \partial_\mu N_\nu -\partial_\nu N_\mu &+[N_\mu, N_\nu]\, , \\     \lambda_{\mu\nu} \equiv  \partial_\mu \lambda_\nu -\partial_\nu
    \lambda_\mu  + i [\lambda_\mu, \lambda_\nu]\, ,\qquad
    &\rho_{\mu\nu} \equiv  \partial_\mu \rho_\nu -\partial_\nu \rho_\mu  +
    i [\rho_\mu, \rho_\nu]\, ,\\ 
    D_\mu l \equiv  \partial_\mu l + i \rho_\mu l - i l \lambda_\mu\, , \qquad
    &D_\mu r \equiv  \partial_\mu r + i \lambda_\mu r - i r \rho_\mu
    \end{aligned}
\end{align}
The contribution of the ghosts has to be added to eq.~\eqref{eq:masterformula}, but can be found with the same techniques.\\
Equation~\eqref{eq:masterformula} is the master formula that we use to find the $1/\epsilon$-poles of the \ewXL\ at one loop in dimensional regularization. However, it can be applied to any theory, assuming that its Lagrangian can be written in the form of eq.~\eqref{eq:lfluct}. The main advantage of this master formula is that all momentum integrals are already performed and the problem of finding the $1/\epsilon$ poles at one loop is reduced to algebraic manipulations of matrices.
\section{The One-Loop Renormalization}
The detailed results are in our publication~\cite{Buchalla:2017jlu}. Here instead, I focus on some of the key aspects of the result. In order to verify our results, we performed several cross checks.
\begin{enumerate}
\item  In the limit
\begin{equation}
  \label{smlimit}
  F(h) = (1+\tfrac{h}{v})^2\, ,\qquad
  V(h) = \frac{m^2_h v^2}{8}\left( -2 (1+\tfrac{h}{v})^2 + (1+\tfrac{h}{v})^4 \right)\, ,\qquad
  M(h) = M_0\, (1+\tfrac{h}{v}),
\end{equation}
the Lagrangian in equation~\eqref{eq:l2} reduces to the Lagrangian of the SM. In this limit, our results reproduce the $\beta$-functions of the SM.
\item  When restricting to loops of Higgs and Goldstone Bosons only, we reproduce the results of~\cite{Guo:2015isa}, where the divergent contributions of these scalar loops were computed.
\item  All five members of our collaboration performed the computation independently, even using different choices of the gauge-fixing Lagrangian in eq.~\eqref{eq:gfix2}.
\end{enumerate}
To find the $\beta$-functions of all the couplings of the Lagrangian, the results of eq.~\eqref{eq:masterformula} have to be projected on an operator basis. At leading order, the operators are defined through eq.~\eqref{eq:l2}. Terms of the form $(\partial_{\mu} h) (\partial^{\mu}h)\, f(h)$ and $\bar\psi i\slashed{D}\psi \, g(h)$, arising from eq.~\eqref{eq:masterformula}, need to be eliminated using field redefinitions~\cite{Buchalla:2013rka}. Further, we have to impose $V'(0)=0$. At next-to-leading order, we project the result of eq.~\eqref{eq:masterformula} to the basis defined in~\cite{Buchalla:2013rka}. The resulting operator structures are sketched in Table~\ref{tab:results}. This computation confirms that
\begin{itemize}
\item the operator basis defined in~\cite{Buchalla:2013rka} is complete, {\it i.e.} all the divergent terms at the one-loop level can be expressed as a sum of the operators of~\cite{Buchalla:2013rka}.  
\item the power counting based on chiral dimensions~\cite{Buchalla:2013eza} gives the right classes of counter terms, {\it i.e.} the operators come with additional factors of weak couplings, as predicted.
\end{itemize}
\newpage
\begin{center}
    \begin{longtable}[!ht]{|c|c|c|c|}
  \hline
  Operator class & Example & \multicolumn{2}{|c|}{Operator count\footnote{The operators are counted with the Higgs functions $\mathcal{F}(h)$ included, meaning different orders in $h$ count to the same operator.}} \\
                 && generated & total~\cite{Buchalla:2013rka}\\
  \hline
  &$ $ &&\\
  $\kappa^{2}UD^{2}h$&$\mathcal{O}_{\beta}=(g'v)^{2} \langle UT_{3}D_{\mu}U^{\dagger}\rangle^{2} \mathcal{F}$ &1&1\\
                 &$$ &&\\
  \hline
  &$ $ &&\\
  $UD^{4}h$&$\mathcal{O}_{D1} =\langle D_{\mu}UD^{\mu}U^{\dagger}\rangle^{2} \, \mathcal{F}$ &5&15\\
  &$ $ &&\\
  \hline
  &$ $ &&\\
  $\kappa UXD^{2}h$&$\mathcal{O}_{XU7} =g' \langle T_{3}D_{\mu}U^{\dagger}D_{\nu}U\rangle B^{\mu\nu} \, \bar{\mathcal{F}}$ &0&8\\
  &$ $ &&\\
  \hline
  &$ $ &&\\
  $\kappa^{2}UX^{2}h$&$\mathcal{O}_{Xh1} =g'^{2} B_{\mu\nu}B^{\mu\nu}\, \bar{\mathcal{F}}$ &0&10\\
  &$ $ &&\\
  \hline
  &$ $ &&\\
  $\kappa^{2}UD\Psi^{2}h$&$\mathcal{O}_{\Psi V1} = i y^{2} (\bar{q}_{L}\gamma^{\mu}q_{L}) \langle UT_{3}D_{\mu}U^{\dagger}\rangle \, \mathcal{F}$ &13&13\\
  &$$ &&\\
  \hline
  &$ $ &&\\
  $\kappa UD^{2}\Psi^{2}h$&$\mathcal{O}_{\Psi S1/2} =y \bar{q}_{L} U P_{\pm}q_{R} \langle D_{\mu}UD^{\mu}U^{\dagger}\rangle \, \mathcal{F} $ &12&30\\
  &$ $ &&\\
  \hline
  &$ $ &&\\
  $\kappa^{2}U\Psi^{2}Xh$&$\mathcal{O}_{\Psi X1/2} =y g' \bar{q}_{L} \sigma_{\mu\nu}U P_{\pm}q_{R} B^{\mu\nu}\, \mathcal{F}$ &0&11\\
  &$  $ &&\\
  \hline
  &$ $ &&\\
  $\kappa^{2}U\Psi^{4}h$&$\mathcal{O}_{LL1}=y^{2} (\bar{q}_{L}\gamma^{\mu}q_{L})(\bar{q}_{L}\gamma_{\mu}q_{L})\, \mathcal{F}$ &22&60\\
  &$ $ &&\\
  \hline
\caption{Schematic summary of the result. The notation of operator classes and operators are as in~\cite{Buchalla:2013rka}, with a generic weak coupling (gauge or Yukawa) $\kappa$.}
\label{tab:results}
\end{longtable}
\end{center}
\section{Conclusions}
Without direct signs of new physics at the LHC, effective field theory techniques have become very popular in recent years. Used in the bottom-up approach, they provide a model-independent framework to look for indirect signs of physics beyond the Standard Model.\\
Here, I introduced the Electroweak Chiral Lagrangian as the most general EFT at the electroweak scale. It generalizes the Higgs couplings of the SM and can therefore also be used to study the properties of the newly-discovered Higgs-like scalar. The Electroweak Chiral Lagrangian applies techniques used in Chiral Perturbation Theory to electroweak physics. The presence of dynamical gauge fields and fermions requires a generalization of the momentum expansion of ChPT to include those fields as well. Since the leading-order Lagrangian is non-renormalizable, the expansion of the EFT is tied to a loop expansion, where NLO counterterms absorb the divergences generated by the leading-order operators. I introduced the concept of chiral dimensions, which are based on the superficial degree of divergence, to keep track of the loop order of a given operator. They provide the desired generalized momentum expansion and can also be applied for example to ChPT coupled to QED~\cite{Urech:1994hd,Knecht:1999ag,Buchalla:2013eza}.\\
Since the power counting is based on the loop structure of the theory, knowing the full divergence structure beyond the superficial degree of divergence is essential for our understanding. Here, I presented our computation~\cite{Buchalla:2017jlu} of the divergence structure of the full Electroweak Chiral Lagrangian. It uses the background-field method and is based on the super-heat-kernel expansion. We find a universal master formula for the $1/\epsilon$-poles in dimensional regularization in eq.~\eqref{eq:masterformula}, which can be applied to any theory with a Lagrangian quadratic in fluctuations of the form of eq.~\eqref{eq:lfluct}, not restricted to the EFT we consider here.\\ 
The result of our computation supports the completeness of the operator basis defined in~\cite{Buchalla:2013rka} and confirms the power counting based on chiral dimensions~\cite{Buchalla:2013eza}, which was based on the superficial degree of divergence.

\bibliography{EWChL}
\bibliographystyle{JHEP}

\end{document}